\documentclass[11pt,a4paper]{article}                                                                    

\input epsf
\epsfverbosetrue
\def\Journal#1#2#3#4{{#1} {\bf #2} (#4) #3}

\def\NPB{{\em Nucl. Phys.} B}
\def\PLB{{\em Phys. Lett.}  B}

\def\PRD{{\em Phys. Rev.} D}
\def\EPC{{\em Eur. Phys. J.} C}
\def\ZPC{{\em Z. Phys.} C}
\def\ZPCe{{\em Z. Phys.-e} C}
\def\CPC{\em Comp. Phys. Comm.}
\def\oos{${\cal O}(\alpha \alpha_s^2)$}
\def\3{\ss}                                                                     
\begin{document}
\title {\begin{flushright}{\large DESY--98--162}  \end{flushright}
\vspace{2cm}
\bf\LARGE  Measurement of Three-jet Distributions in Photoproduction at HERA \\
\vspace{1cm}}
                    
\author{ZEUS Collaboration}
\date{}

\maketitle
\begin{abstract}
\noindent
The cross section for the photoproduction of events containing 
three jets with a three-jet invariant mass of 
$M_{\mbox{\scriptsize 3J}} > 50$~GeV has been measured with the ZEUS detector
at HERA.  The three-jet angular
distributions are
inconsistent with a uniform population of the available phase 
space but are well described by
parton shower models and ${\cal O}(\alpha \alpha_s^2)$
pQCD calculations.
Comparisons with the parton shower model indicate a strong 
contribution
from initial state radiation as well as a sensitivity
to the effects of colour coherence.

\end{abstract}

\pagestyle{plain}
\thispagestyle{empty}
\clearpage
%
%
%
%
%

\pagenumbering{Roman}                                                                              
                                                   %
\begin{center}                                                                                     
{                      \Large  The ZEUS Collaboration              }                               
\end{center}                                                                                       
  J.~Breitweg,                                                                                     
  S.~Chekanov,                                                                                     
  M.~Derrick,                                                                                      
  D.~Krakauer,                                                                                     
  S.~Magill,                                                                                       
  B.~Musgrave,                                                                                     
  J.~Repond,                                                                                       
  R.~Stanek,                                                                                       
  R.~Yoshida\\                                                                                     
 {\it Argonne National Laboratory, Argonne, IL, USA}~$^{p}$                                        
\par \filbreak                                                                                     
  M.C.K.~Mattingly \\                                                                              
 {\it Andrews University, Berrien Springs, MI, USA}
% ...
\newpage                                                                                           
% ...
$^{\    3}$ supported by Worldlab, Lausanne, Switzerland \\                                        
% ...
$^{  32}$ partially supported by the Foundation for German-Russian Collaboration                   
DFG-RFBR \\ \hspace*{3.5mm} (grant no. 436 RUS 113/248/3 and no. 436 RUS 113/248/2)\\              
% ...
                                                           %
                                                           %
\newpage   
                                                           %
                                                           %
\begin{tabular}[h]{rp{14cm}}                                                                       
$^{a}$ &  supported by the Natural Sciences and Engineering Research                               
          Council of Canada (NSERC)  \\                                                            
$^{b}$ &  supported by the FCAR of Qu\'ebec, Canada  \\                                            
$^{c}$ &  supported by the German Federal Ministry for Education and                               
          Science, Research and Technology (BMBF), under contract                                  
          numbers 057BN19P, 057FR19P, 057HH19P, 057HH29P, 057SI75I \\                              
$^{d}$ &  supported by the MINERVA Gesellschaft f\"ur Forschung GmbH, the                          
German Israeli Foundation, and by the Israel Ministry of Science \\                                
$^{e}$ &  supported by the German-Israeli Foundation, the Israel Science                           
          Foundation, the U.S.-Israel Binational Science Foundation, and by                        
          the Israel Ministry of Science \\                                                        
$^{f}$ &  supported by the Italian National Institute for Nuclear Physics                          
          (INFN) \\                                                                                
$^{g}$ &  supported by the Japanese Ministry of Education, Science and                             
          Culture (the Monbusho) and its grants for Scientific Research \\                         
$^{h}$ &  supported by the Korean Ministry of Education and Korea Science                          
          and Engineering Foundation  \\                                                           
$^{i}$ &  supported by the Netherlands Foundation for Research on                                  
          Matter (FOM) \\                                                                          
$^{j}$ &  supported by the Polish State Committee for Scientific Research,                         
          grant No. 115/E-343/SPUB/P03/002/97, 2P03B10512,                                         
          2P03B10612, 2P03B14212, 2P03B10412, 2P03B05315 \\                                        
$^{k}$ &  supported by the Polish State Committee for Scientific                                   
          Research (grant No. 2P03B08614) and Foundation for                                       
          Polish-German Collaboration  \\                                                          
$^{l}$ &  partially supported by the German Federal Ministry for                                   
          Education and Science, Research and Technology (BMBF)  \\                                
$^{m}$ &  supported by the Fund for Fundamental Research of Russian Ministry                       
          for Science and Edu\-cation and by the German Federal Ministry for                       
          Education and Science, Research and Technology (BMBF) \\                                 
$^{n}$ &  supported by the Spanish Ministry of Education                                           
          and Science through funds provided by CICYT \\                                           
$^{o}$ &  supported by the Particle Physics and                                                    
          Astronomy Research Council \\                                                            
$^{p}$ &  supported by the US Department of Energy \\                                              
$^{q}$ &  supported by the US National Science Foundation \\                                       
\end{tabular}                                                                                      
                                                           %

                                               %
\newpage
\setcounter{page}{1}
\pagenumbering{arabic}
\section{Introduction}
Calculations of photoproduction processes beyond leading order in
perturbative QCD (pQCD) predict a rich variety of phenoma.  Some of
these can be studied in final states containing more than two jets.
Also, the study of multijet production provides sensitive tests
of extensions to fixed order theories such as parton shower models.
The properties of multijet events in hadronic collisions have been 
the subject of earlier studies~\cite{UA2_86,D0_96,CDF_96}.
Dijet photoproduction accompanied by a third, low transverse energy
cluster has been studied by ZEUS~\cite{remnant_95}.
In this paper, 
cross sections and angular distributions for three or more 
moderately high 
transverse energy jets in photoproduction are presented for the 
first time.

Apart from the azimuthal orientation, a system of two massless jets 
can be completely specified in its centre-of-mass (CM) frame 
by the two-jet invariant mass, $M_{\mbox{\scriptsize 2J}}$ and 
$\cos \vartheta^*$, where $\vartheta^*$ is the angle between
the jet axis and the beam-line.  The distribution in $\cos \vartheta^*$
for photoproduction of dijets is forward-backward peaked
with sensitivity to the spin of the
exchanged fermion or boson~\cite{dijet_96}.

\begin{figure}[p]
\centering
\leavevmode
\epsfxsize=7.0cm.\epsfbox{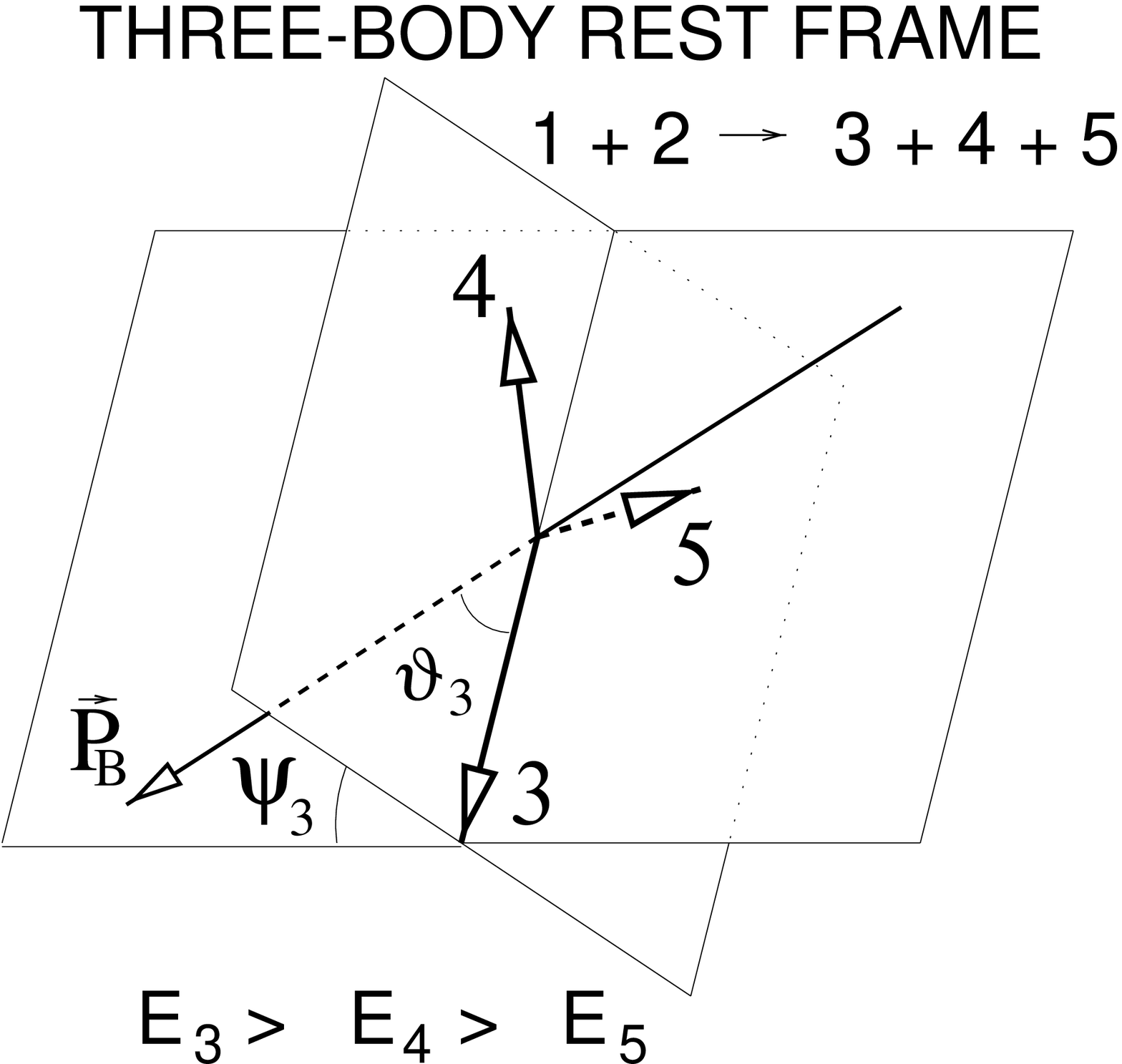}
\caption{Illustration of the angles $\vartheta_3$ and $\psi_3$ for a particular
three-jet configuration. The beam direction is indicated by $\vec{p}_B$.\label{fig:diagram}}
\end{figure}

A set of observables describing events with an arbitrary number of 
jets has been proposed which spans the multijet parameter space, 
facilitates the interpretation of the data within pQCD and reduces to 
$M_{\mbox{\scriptsize 2J}}$ and $\cos \vartheta^*$ for the dijet 
case~\cite{Geer_96}.  
For three massless jets there are five parameters which are defined
in terms of the energies, $E_i$, and momentum three-vectors, 
$\vec{p}_i$, of the jets in the three-jet CM frame and 
$\vec{p}_{B}$, the 
beam direction.\footnote{We take the nominal beam direction as
$\vec{p}_B = \hat{z}$.  In the ZEUS coordinate
system the $z$-axis is defined to be in the proton beam direction.  
Polar angles, $\vartheta$, are measured with respect to the $z$-axis
and pseudorapidity is defined as 
$\eta = - \ln(\tan \frac{\vartheta}{2})$.}
The jets are numbered, 3, 4 and 5 in order 
of decreasing energy as illustrated in the schematic 
drawing, Fig.~\ref{fig:diagram}.
The parameters are: the three-jet invariant 
mass, $M_{\mbox{\scriptsize 3J}}$;
the energy-sharing quantities $X_3$ and $X_4$,
\begin{equation}
X_i \equiv \frac{2E_i}{M_{\mbox{\scriptsize 3J}}};
\end{equation}
the cosine of the scattering angle of the highest energy jet with
respect to the beam,
\begin{equation}
\cos \vartheta_3 \equiv \frac{\vec{p}_{B} \cdot 
                      \vec{p}_3}{|\vec{p}_{B}| |\vec{p}_3|};
\end{equation}
and $\psi_3$, the angle between 
the plane containing the highest energy jet 
and the beam and 
the plane containing the three jets.  The latter is defined by
\begin{equation}  
\cos{\psi_3} \equiv \frac{(\vec{p}_3 \times \vec{p}_{B}) \cdot 
                              (\vec{p}_4 \times \vec{p}_5)}
                              {|\vec{p}_3 \times \vec{p}_{B}| 
                               |\vec{p}_4 \times \vec{p}_5|}.
\end{equation}

The definition of the angles $\vartheta_3$ and $\psi_3$ is illustrated in 
Fig.~\ref{fig:diagram}.  Since $\vartheta_3$ involves only the highest
energy jet, the distribution of $\cos \vartheta_3$ in three-jet processes
may be expected to follow closely the distribution of
$\cos \vartheta^*$ in dijet events.  The $\psi_3$ angle, on the other 
hand, reflects the orientation of the lowest energy jet.  In the 
case where this jet arises from initial-state 
radiation, the coherence property of QCD will tend to orient the 
third jet close to the incoming proton or photon direction.
The two planes shown in Fig.~1 will therefore 
tend to coincide
leading to a $\psi_3$ distribution which peaks toward 0 and $\pi$.

This paper presents the three-jet inclusive cross section in 
photoproduction and the distribution of the three-jet events with 
respect to $M_{\mbox{\scriptsize 3J}}$, $X_3$, $X_4$, $\cos \vartheta_3$ and $\psi_3$.
This work
was performed with the ZEUS detector using 16~pb$^{-1}$ of data
delivered by HERA in 1995 and 1996.

\section{Experimental Conditions}
In this period HERA operated with protons of energy
$E_p = 820$~GeV and positrons of energy $E_e = 27.5$~GeV.  The
ZEUS detector is described in detail in~\cite{ZEUS_1,ZEUS_2}.
The main components used in the present analysis are the central
tracking system positioned in a 1.43~T solenoidal magnetic field
and the uranium-scintillator sampling calorimeter (CAL).  The
tracking system was used to establish an interaction vertex.
Energy deposits in the CAL were used in the jet finding and to measure
jet energies.  The CAL is hermetic and consists of 5918 cells each
read out by two photomultiplier tubes.  Under test beam conditions
the CAL has energy resolutions of $18\%/\sqrt{E\mbox{ (GeV)}}$ for electrons and
$35\%/\sqrt{E\mbox{ (GeV)}}$ for hadrons.  Jet energies were 
corrected for the 
energy lost in inactive material in front of the CAL which is typically
about one radiation length (see Section~\ref{sect:corr}).  The effects of 
uranium noise were minimized by discarding cells in the electromagnetic
or hadronic sections if they had energy deposits of less than
60~MeV or 110~MeV, respectively.
The luminosity was measured from the rate of the bremsstrahlung
process $e^+ p \rightarrow e^+ p \gamma$.
A three-level trigger was used to
select events online~\cite{ZEUS_2, dijet_98}.

\section{Analysis}
\subsection{Offline Cleaning Cuts}
To reject residual beam-gas and
cosmic ray backgrounds, tighter cuts using the final $z$-vertex
position, other tracking information and timing information are
applied offline.
Neutral current deep inelastic scattering (DIS) events with an 
identified scattered positron candidate in the CAL are removed 
from the  sample as described in detail 
elsewhere~\cite{dijet_96,dijet_98}.
Charged current DIS events are rejected by a cut on the missing 
transverse  momentum measured in the CAL.
Finally, a restriction is made on the range of $y$, the fraction of the
positron's energy carried by the incoming photon.
The requirement, $0.15 < y_{\mbox{\scriptsize JB}} < 0.65$ is made
where $y_{\mbox{\scriptsize JB}}$ is an estimator of $y$ which is
determined from the CAL energy deposits according to the
Jacquet-Blondel method~\cite{YJB}.
This requirement corresponds to accepting events in
the range $0.2 < y < 0.8$.
These cuts restrict photon virtualities to less than 
about 1~GeV$^2$
with a median of around $10^{-3}$~GeV$^2$.

\subsection{Jet Finding}
Jets are found using the KTCLUS~\cite{Mike_93} finder in the
inclusive mode~\cite{Ellis_93}.  This is a clustering algorithm 
which combines objects with small relative transverse energy 
into jets.  It is invariant under Lorentz boosts along the beam axis
and is ideal for the study of multijet processes
since it suffers from no 
ambiguities due to overlapping jets.
Once the jets are determined, their
transverse energy, pseudorapidity and azimuth
are calculated according to the Snowmass 
convention~\cite{snow};
$E_T^{\mbox{\scriptsize jet}} = \sum_i E_{T_i}$,
$\eta^{\mbox{\scriptsize jet}} = (1/E_T^{\mbox{\scriptsize jet}})\sum_i E_{T_i} \eta_i$ and 
$\varphi^{\mbox{\scriptsize jet}} = (1/E_T^{\mbox{\scriptsize jet}})\sum_i E_{T_i} \varphi_i$,
where the sum runs over all objects assigned to the jet.
The energies and
three-vectors of the jets are then determined from the 
$E_T^{\mbox{\scriptsize jet}}$, $\eta^{\mbox{\scriptsize jet}}$ and 
$\varphi^{\mbox{\scriptsize jet}}$.

The objects input to the jet algorithm may be hadrons in
a simulated hadronic final
state, the final state partons of a pQCD calculation, or 
energy deposits in the detector.
In the following, a jet quantity constructed from CAL cells with no 
energy correction has the superscript ``CAL'' while a jet quantity 
constructed from CAL cells and then subjected to a correction for 
energy loss in inactive material has the superscript ``COR''. 
There is no additional superscript 
for quantities referring to jets of final state partons or hadrons.

\subsection{Monte Carlo Event Simulation}
The response of the detector to jets and the acceptance and
smearing of the measured distributions are determined using
samples of events generated from Monte Carlo (MC) simulations.
We have used the programs 
{\sc PYTHIA~5.7}~\cite{PYTHIA} and {\sc HERWIG~5.9}~\cite{HERWIG} 
which implement the leading order matrix elements followed by
parton showers.  In these simulations multijet events can
originate through this parton shower mechanism.
Colour coherence in the parton shower is treated
differently in the two models.
In PYTHIA, parton showers are evolved in the squared mass of the
branching parton with colour coherence effects implemented as a 
restriction on the opening angle of the radiation.  In contrast,
in HERWIG a parton shower evolution variable is chosen which
automatically limits the branching to an angular ordered region.
For both models, leading order direct and resolved processes are 
generated separately and combined according to the ratio of
their generated cross sections.
For the
uncorrected distributions presented in this section the 
minimum transverse momentum of the partonic hard scatter
($\hat{p}_{T}^{\mbox{\scriptsize min}}$) was set to 4~GeV.  In
Section~\ref{sect:results}, corrected cross sections are presented
and compared with the predictions from HERWIG and PYTHIA with
$\hat{p}_{T}^{\mbox{\scriptsize min}}=8$~GeV (our conclusions are
insensitive to this parameter).
The photon parton densities
used were GRV~LO~\cite{GRV}
 and the proton parton densities were CTEQ4~LO~\cite{CTEQ}.
In the HERWIG simulation of the resolved processes, multiparton 
interactions have been included (these are not important in this
kinematic regime, as discussed in Section~\ref{sect:results}).

\begin{figure}[p]
\centering
\leavevmode
\epsfxsize=13.cm\epsfbox{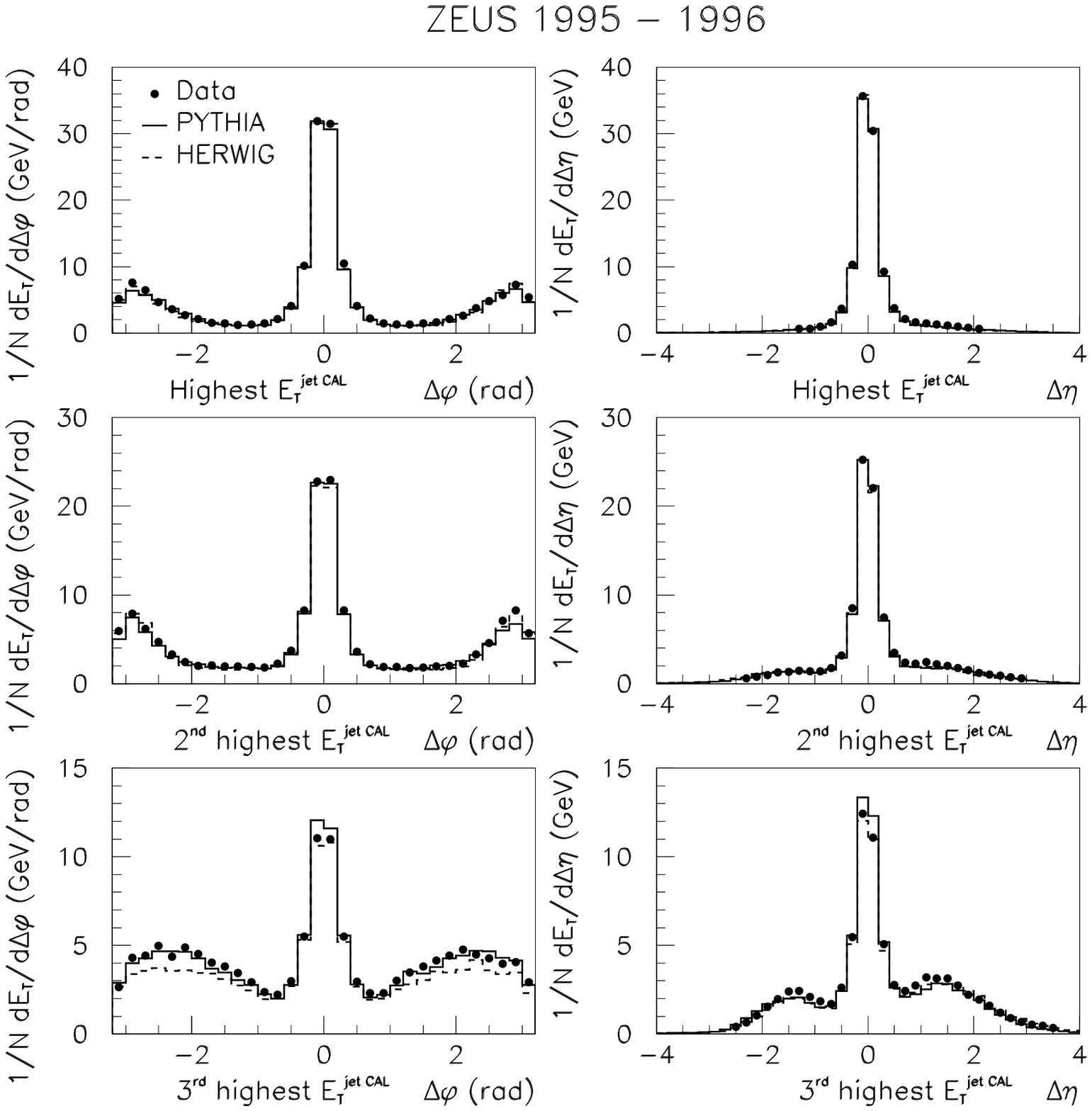}
\caption{Uncorrected transverse energy flow with respect to
the jet axis in the laboratory frame for three-jet events
in order of $E_T^{\mbox{\scriptsize jet CAL}}$.
On the left the uncorrected energy flow with respect to $\varphi$ is shown
for cells within one unit of $\eta$ of the jet axis while on
the right the profile with respect to $\eta$ is shown for cells
within one radian of $\varphi$ of the jet axis.  The data are
shown as black dots while the PYTHIA and HERWIG predictions are
shown by the solid and dashed histograms, respectively.}
\label{fig:profiles}
\end{figure}

The quality of these simulations
is illustrated in Fig.~\ref{fig:profiles}
which
shows the transverse energy flow around the jet axes.
In this comparison the events have 
two jets with  
$E_T^{\mbox{\scriptsize jet CAL}} > 5$~GeV and a third jet with
$E_T^{\mbox{\scriptsize jet CAL}} > 4$~GeV and 
jet pseudorapidities  
$|\eta^{\mbox{\scriptsize jet CAL}}| < 2.4$.  The additional 
requirements on the CM quantities,
$M_{\mbox{\scriptsize 3J}}^{\mbox{\scriptsize CAL}} > 42$~GeV, 
$|\cos \vartheta_3|^{\mbox{\scriptsize CAL}} < 0.8$ and 
$X_3^{\mbox{\scriptsize CAL}} < 0.95$,
have also been applied.
These conditions represent those on the selected events 
(described in Section~\ref{sect:selection}) to a good 
approximation.
The jets have a narrow core with little transverse energy
in the pedestal, except for the lowest 
$E_T^{\mbox{\scriptsize jet CAL}}$ jet where significant contribution
to the ``pedestal'' from the other two jets in the event would
be expected.  In the $\Delta \varphi$ profiles of the two highest
$E_T^{\mbox{\scriptsize jet CAL}}$ jets, the peaks near $\pm \pi$
indicate that these are roughly back-to-back.
The PYTHIA and HERWIG event samples were
passed through a detailed simulation of the ZEUS detector and
the same selection criteria as for the data were applied.  The MC
samples provide a reasonable description of
the energy flow in these three-jet events.  HERWIG
generates somewhat too little transverse energy in regions
far from the jet core in $\varphi$ for the lowest
transverse energy jet, while PYTHIA slightly overestimates the 
transverse energy in the core of this jet.  These
models are also able to reproduce satisfactorily the 
$y_{\mbox{\scriptsize JB}}$
distribution for the three-jet events, the lab-frame 
$E_T^{\mbox{\scriptsize jet CAL}}$ and 
$\eta^{\mbox{\scriptsize jet CAL}}$ distributions as well as
the transverse and longitudinal components of the boost from the lab-frame to the 
CM-frame (not shown).

\subsection{Jet Energy Corrections}\label{sect:corr}
Jets of hadrons lose about 15\% of their transverse energy when 
passing through inactive material before impinging on the 
CAL.  This
energy loss has been corrected using the MC
samples~\cite{dijet_98,incjet_98}.  The KTCLUS algorithm was
applied to the hadronic final state and from comparison of
these hadron jets with the CAL jets obtained after the
detector simulation, correction factors were determined
as a function of
$E_T^{\mbox{\scriptsize jet CAL}}$ and 
$\eta^{\mbox{\scriptsize jet CAL}}$.
After applying these corrections the average shift in 
$E_T^{\mbox{\scriptsize jet}}$ within the MC simulation is 
less than 2\% and the
$E_T^{\mbox{\scriptsize jet}}$ resolution
is 14\%.  
This may be compared with the global jet energy scale uncertainty of
$\pm 5$\%~\cite{dijet_98}.
The correction also reduces the shift in the
reconstruction of $M_{\mbox{\scriptsize 3J}}$ from 16\% to less than 1\%.  After
these corrections for jet energy loss the average resolutions are
8\% in $M_{\mbox{\scriptsize 3J}}$, 
0.03 units in $X_3$, 
0.05 units in $X_4$, 
0.03 units in $\cos \vartheta_3$ and 0.1 radians in $\psi_3$ and
the distributions are well centred on their expected values.

\subsection{Event Selection}\label{sect:selection}
After the jet energy correction the events are required to have
at least two
jets with $E_T^{\mbox{\scriptsize jet COR}} > 6$~GeV, a third 
jet with
$E_T^{\mbox{\scriptsize jet COR}} > 5$~GeV and jet pseudorapidities 
in the range
$|\eta^{\mbox{\scriptsize jet COR}}| < 2.4$.
The requirement of high transverse energy for the jets ensures that the 
process should be calculable within pQCD.  However, it introduces a 
bias in the angular distributions by excluding jets that are produced close 
to the beam-line.  We make the additional requirements
$M_{\mbox{\scriptsize 3J}}^{\mbox{\scriptsize COR}} > 50$~GeV, 
$|\cos \vartheta_3|^{\mbox{\scriptsize COR}} < 0.8$ and 
$X_3^{\mbox{\scriptsize COR}} < 0.95$ to minimize such a bias.
After these cuts the mean transverse energy of the highest,
second-highest and third-highest transverse energy jet is about
20~GeV, 15~GeV and 10~GeV.
From 16~pb$^{-1}$ of data, 2821
events are selected.  Around 15\% of these
have a fourth jet with 
$E_T^{\mbox{\scriptsize jet COR}} > 5$~GeV,
in agreement with the prediction of the parton shower models.
Backgrounds from beam gas and cosmic ray 
events, determined from unpaired bunch crossings, are
negligible.  The DIS contamination, determined using
Monte Carlo techniques, is around 1\% and neglected.

\subsection{Acceptance Correction}
The MC samples have been
used to correct the data for the inefficiencies of the trigger
and the offline selection cuts and for migrations caused by detector 
effects.
The correction factors are calculated as the ratio
$N_{\mbox{\scriptsize true}} / N_{\mbox{\scriptsize rec}}$
in each measured bin where $N_{\mbox{\scriptsize true}}$ is the 
number of events
generated in the bin and $N_{\mbox{\scriptsize rec}}$ is the
number of events reconstructed in the bin after detector smearing
and all experimental cuts.  The final bin-by-bin correction factors
lie between about 0.7 and 1.3, the dominant effect arising from
migrations across the $M_{\mbox{\scriptsize 3J}}$ threshold.  The cross sections were 
determined using the corrections obtained with PYTHIA.

\subsection{Systematic Uncertainties}
A detailed study of the sources contributing to the systematic 
uncertainties of the measurements was performed~\cite{Esther}.
Only the significant sources are listed here.
\begin{itemize}
   \item The acceptance correction was performed using HERWIG 
    instead of PYTHIA.  The uncertainties associated with the model
    are typically around 20\% and this forms the dominant 
    uncertainty on the area-normalized distributions.
   \item The absolute energy scale of the detector response to
    jets with $E_T^{\mbox{\scriptsize jet}} > 5$~GeV is known to
    $\pm 5$\%~\cite{dijet_98}.
    This leads to an uncertainty of 
    15 to 20\% on the
    cross section.  This is the dominant systematic uncertainty
    on the normalization of the cross section but as this 
    uncertainty is highly
    correlated between bins it has a negligible affect on the
    area-normalized distributions.
   \item The results were recalculated 
    allowing for fluctuations from outside the selected 
    kinematic region by relaxing each of the
    cut parameters by 1~$\sigma$ of the resolution.  This effect
    is typically 5\%.
\end{itemize}
The systematic uncertainties have been added in quadrature to the 
statistical errors and this is shown as the outer error bars in the 
figures, with the exception of the absolute jet energy scale 
uncertainty which
is shown as a shaded band for the cross sections.  An
overall normalization uncertainty of 1.5\% from the luminosity 
determination has not been included.

\section{Results and Discussion\label{sect:results}}

The three-jet inclusive cross section is presented
for events having at least two jets with 
$E_T^{\mbox{\scriptsize jet}} > 6$~GeV and a third jet with
$E_T^{\mbox{\scriptsize jet}} > 5$~GeV
where the jets satisfy $|\eta^{\mbox{\scriptsize jet}}| < 2.4$.
This cross section refers to jets in the hadronic final state.
To minimize the effects of these jet cuts on the distributions
of physical interest, the requirements
$M_{\mbox{\scriptsize 3J}} > 50$~GeV, $|\cos \vartheta_3| < 0.8$ and $X_3 < 0.95$ have
been imposed.
The cross section is presented for photon-proton CM energies 
$W_{\gamma p}$ in the range 134~GeV$ < W_{\gamma p} < 269$~GeV
and the negative square of the invariant mass of the incoming
photon extending to 1~GeV$^2$.
The cross section is
$\sigma = 162 \pm 4(\mbox{stat.}) 
                  ^{+ 16} _{- 6}(\mbox{sys.})
                  ^{+ 32} _{- 25}(\mbox{energy scale})$~pb.

A study using the PYTHIA MC indicates that 
hadronization effects are small ($\sim 5$\%), and 
flat in the distributions presented here~\cite{Esther}.
The measurements are directly confronted with \oos\ pQCD 
calculations from two groups of authors~\cite{Harris,Klasen_1,Klasen_2}.
The CTEQ4~LO~\cite{CTEQ} proton parton densities 
and the GRV~LO~\cite{GRV} photon parton densities have been used in these
calculations.
The renormalization and factorization scales, $\mu$, have
been chosen to equal $E_T^{\mbox{\scriptsize max}}$, where 
$E_T^{\mbox{\scriptsize max}}$ is the largest of the 
$E_T^{\mbox{\scriptsize jet}}$ values of the three jets.
$\alpha_s$ was calculated at one loop with
$\Lambda^{(5)}_{\overline{\mbox{\scriptsize MS}}} = 181$~MeV.
As the calculations are leading order for three-jet production the
normalization uncertainty due to the choice of $\mu$ is expected to be
large.  An uncertainty of a factor of two in the cross section for variation of 
$\mu$ between $E_T^{\mbox{\scriptsize max}} / 2$ and 
$2 E_T^{\mbox{\scriptsize max}}$ has been 
quoted~\cite{Klasen_2}.

\begin{figure}[p]
\centering
\leavevmode
\epsfxsize=13.cm.\epsfbox{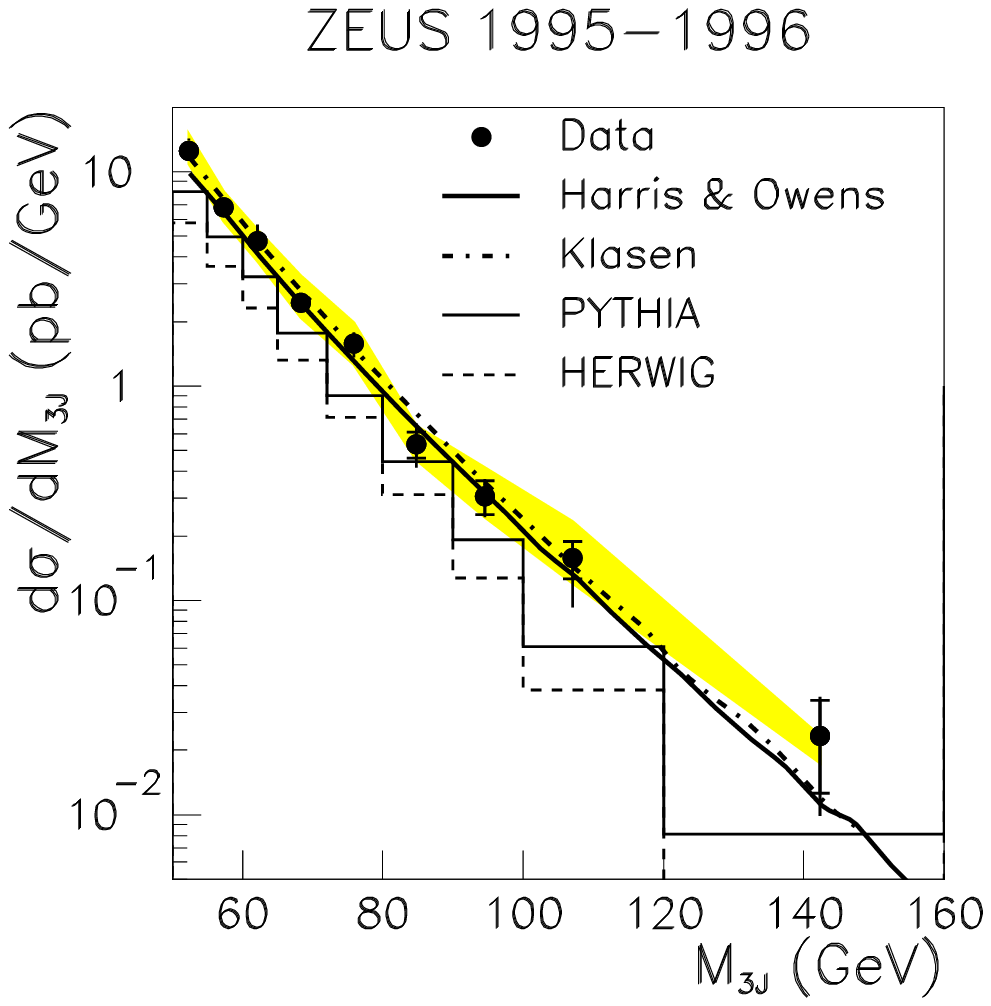}
\caption{The measured three-jet cross section with respect to 
the three-jet invariant 
mass, $d\sigma/d M_{\mbox{\scriptsize 3J}}$, is shown by the black dots where the inner 
error bar shows the statistical error and the outer error bar is the 
sum in quadrature of the statistical error and the systematic 
uncertainty.  The jet energy-scale uncertainty, which is highly correlated 
between bins, is shown separately as the shaded band.  \oos\ pQCD 
calculations by Harris \& Owens and Klasen are shown by the thick solid 
and dot-dashed lines, respectively.  The thin solid and dashed histograms show the 
predictions from two different parton shower 
models, PYTHIA and HERWIG.\label{fig:m3j}} 
\end{figure}

The three-jet invariant mass distribution is shown in 
Fig.~\ref{fig:m3j}.
The cross section falls approximately exponentially from the threshold
value at 50~GeV to the highest measured value, around 150~GeV.
The data are compared with the two \oos\ pQCD calculations.
These are in good agreement with the data, even though
the calculations are leading order for this process.
The $M_{\mbox{\scriptsize 3J}}$ distributions predicted by the parton shower models
PYTHIA and HERWIG
are also in agreement with the data in shape although the predicted  
cross sections are too low by 30-40\%.

\begin{figure}[p]
\centering
\leavevmode
\epsfxsize=13.cm.\epsfbox{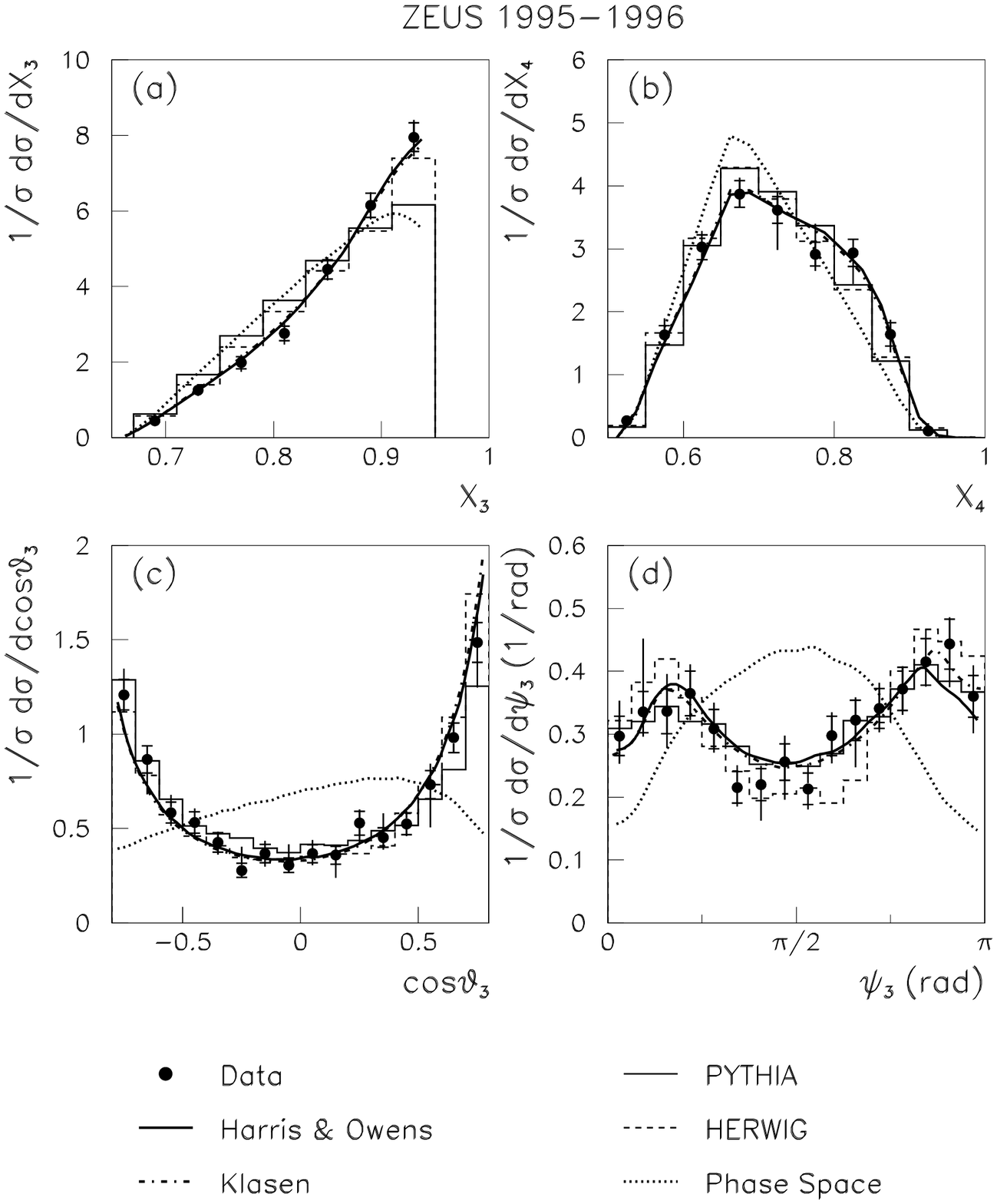}
\caption{The distributions of the energy sharing quantities, $X_3$ and 
$X_4$, are shown by the black dots in (a) and (b), respectively, and the
distributions of the $\cos \vartheta_3$ and $\psi_3$ are shown
in (c) and (d).  Inner error bars show the statistical error and the 
outer error bars show the quadratic sum of this with the
systematic uncertainty.  The fixed-order pQCD predictions are shown by the
thick solid and dot-dashed lines and the parton shower model predictions 
are shown by the thin solid and dashed histograms.  The phase space 
distribution of three jets is indicated by the dotted 
line.\label{fig:distributions}}
\end{figure}

The distributions of the fraction of the available energy taken by 
the highest and second-highest energy jets are shown in 
Figs.~\ref{fig:distributions} (a) 
and (b), respectively.
Here the prediction for three jets uniformly distributed in the 
available phase space (i.e. with a constant matrix element) is also shown
 as the dotted curve.
The parton shower models give a reasonable
description of these energy sharing quantities.
The pQCD calculations (overlapping)
are in excellent agreement with these distributions.  However, the
similarity between the measured distributions and the three-body phase 
space prediction indicates that these distributions have little 
sensitivity to the pQCD matrix elements.

In Figs.~\ref{fig:distributions} (c) and (d) the 
$\cos \vartheta_3$ and $\psi_3$ distributions are shown.
These angular distributions are dramatically different from the 
distributions obtained from phase space, demonstrating
that these quantities are sensitive to the pQCD 
matrix elements.
The $\cos \vartheta_3$ distribution has forward and backward
peaks, as expected.
The \oos\ pQCD as well as the parton shower
calculations, which take into account the dependence of this 
distribution on the spin of the exchanged quark or gluon,
are in good agreement with the data.
The jet algorithm and minimum 
$E_T^{\mbox{\scriptsize jet}}$
requirements deplete the data near $\psi_3 \sim 0$ 
and $\pi$ as indicated by the shape of the phase space curve
in Fig.~\ref{fig:distributions}(d).  
With this taken into account, the data indicate a strong tendency for the three-jet plane to
lie near the plane containing the beam and the highest energy jet.
This effect is reproduced in the \oos\ matrix element calculations.
It is interesting that the parton shower Monte Carlo programs PYTHIA 
and HERWIG are also able to provide a reasonable
representation of the shape of the $\psi_3$ distribution.

Including a simulation of multiparton interactions in the parton 
shower programs has been found to improve significantly 
the description of low $E_T^{\mbox{\scriptsize jet}}$ 
photoproduction~\cite{dijet_98}.
In the present study the sensitivity to multiparton interactions has 
been investigated using 
both PYTHIA and HERWIG~\cite{Esther}.  In neither case do secondary 
parton interactions cause a significant difference in the 
three-jet cross section in this kinematic regime, or in the angular 
distributions generated.
It appears therefore that the third jet arises here from the parton 
shower and is not due to a second hard scatter.

\begin{figure}[p]
\centering
\leavevmode
\epsfxsize=13.cm.\epsfbox{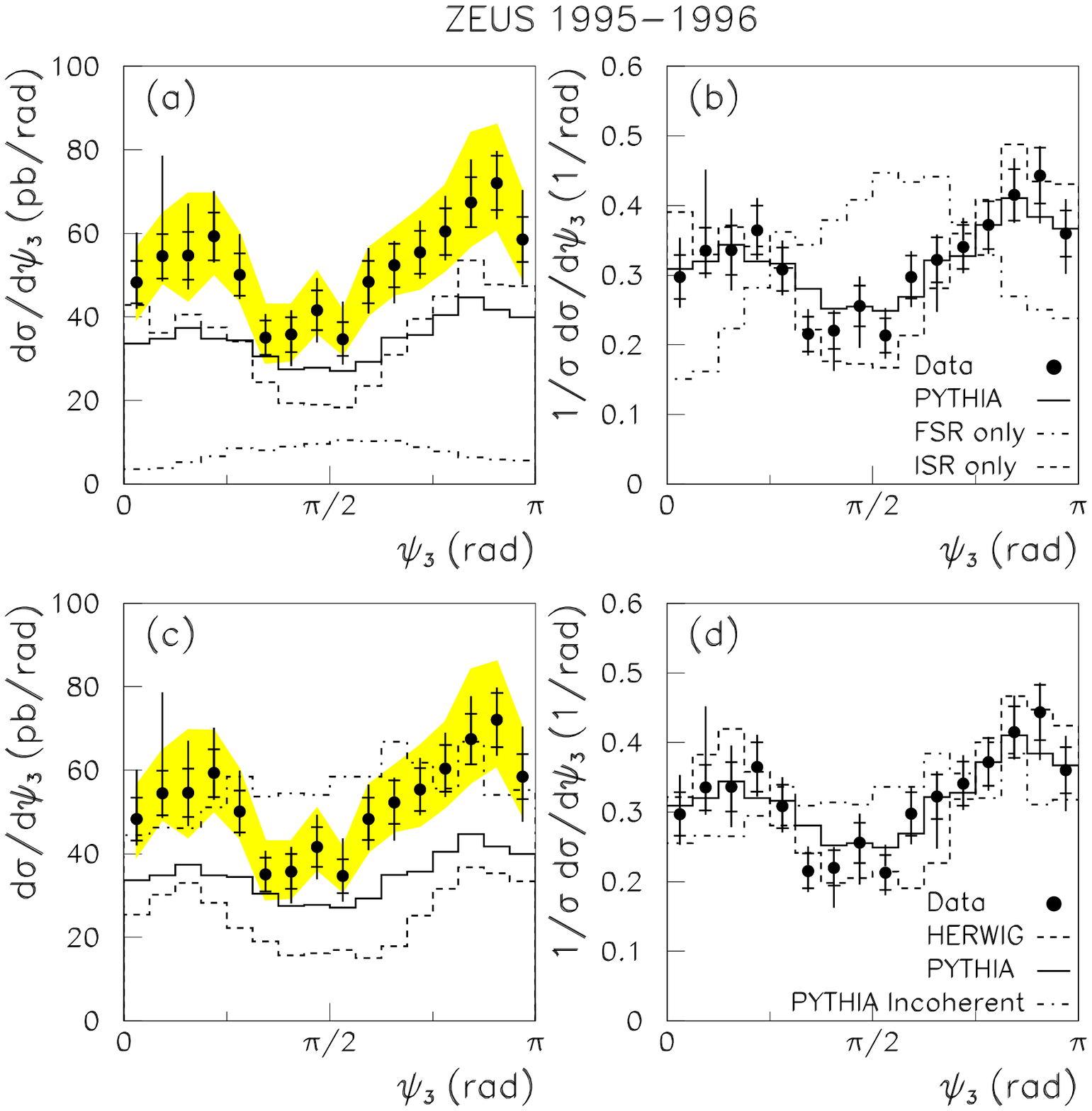}
\caption{The measured cross section 
$d\sigma / d\psi_3$ is shown in (a) and (c) 
and the area-normalized distribution of $\psi_3$ is
shown in (b) and (d).  The error bars are as described previously with the correlated 
systematic uncertainty due to the jet energy-scale shown as the shaded band 
in (a) and (c).  
The solid histogram shows the default PYTHIA prediction.
In (a) and (b) the dashed and dot-dashed histograms show the 
predictions
from PYTHIA with final state radiation switched off and with initial
state radiation switched off.
In (c) and (d) the dashed and dot-dashed 
histograms show the predictions
of HERWIG and of PYTHIA with colour coherence 
switched off.
\label{fig:psis}}
\end{figure}

Within the parton shower prescription, it is possible to separate the
contributions of 
initial and final state parton showers.
In Fig.~\ref{fig:psis}(a) the 
three-jet cross section as a function of $\psi_3$ is shown and 
compared with the predictions of PYTHIA.  The MC events have
been separated into three samples; initial-state 
radiation only, final-state radiation only, and 
default PYTHIA which includes the interference of these two.  
The area-normalized distributions
of $\psi_3$ are compared with these models in Fig.~\ref{fig:psis}(b).
Both the normalization and shape of these distributions indicate
that the observed three-jet production occurs predominantly through initial
state radiation with the final state radiation making a small
contribution.

The QCD phenomenon of colour coherence is implemented in 
the PYTHIA parton shower model by prohibiting radiation into 
certain angular regions which are determined by the colour flow 
of the primary scatter.
It is possible within this model to switch QCD colour coherence
on and off.
Figs.~\ref{fig:psis}(c) and (d) again show the cross section 
and the area-normalized distribution 
with respect to $\psi_3$ compared with 
the HERWIG and PYTHIA predictions.
The data lie above these predictions as previously mentioned, however
this discrepancy is not regarded as significant in view of the limited
order of the calculation.  The predictions do reproduce reasonably
well the shape of the $\psi_3$ distribution.
This is not the case if the simulation is done with 
colour coherence switched off.  The incoherent PYTHIA prediction 
is relatively flat in $\psi_3$.  Coherence 
reduces the phase space available for large angle emissions as indicated
by the drop in cross section around $\psi_3 \sim \pi / 2$ for default 
PYTHIA.
Colour coherence in the parton shower is needed to 
describe the shape of this distribution.
QCD colour coherence seems to be a stronger effect in HERWIG than
in PYTHIA however
the present data are not precise enough to discriminate between these
two simulations.

\section{Summary}
The inclusive cross section for the photoproduction of three 
jets has been measured by the ZEUS collaboration at HERA.  \oos\ pQCD 
calculations are able to describe the cross section 
$d\sigma / d M_{\mbox{\scriptsize 3J}}$, 
while parton shower models underestimate the cross section but are 
consistent in shape with the $M_{\mbox{\scriptsize 3J}}$ distribution.  The angular 
distributions of the three jets are inconsistent with a uniform population 
of the available phase space but are well described by both fixed-order 
pQCD calculations and 
parton shower Monte Carlo models.  Simulation of multiparton interactions
does not help to describe the data in this kinematic regime.
Within the parton shower model the three-jet events are found to occur 
predominantly due to initial state radiation, and the fundamental QCD 
phenomenon of colour coherence is seen to have an important effect on 
the angular distribution of the third jet.

\section*{Acknowledgements}
The strong support and encouragement of the DESY Directorate have
been invaluable, and we are much indebted to the HERA machine group
for their inventiveness and diligent efforts.  The design,
construction and installation of the ZEUS detector have been made
possible by the ingenuity and dedicated efforts of many people from 
inside DESY and from the home institutes who are not listed as authors.
Their contributions are acknowledged with great appreciation.
We warmly thank B.~Harris, M.~Klasen and J.~Owens for providing their 
theoretical calculations.

\end{document}